# Nonlinear propagation effect in x-ray parametric amplification during high harmonic generation

J. SERES,[1]* E. SERES,[1] C. SERRAT,[2] T. H. DINH,[3] N. HASEGAWA,[3] M. ISHINO,[3] M. NISHIKINO,[3] K. NAKANO,[4] AND S. NAMBA[4]

[1]*Institute of Atomic and Subatomic Physics E-141, Vienna University of Technology, Stadionalle 2, 1020 Vienna, Austria*
[2]*Department of Physics, Polytechnic University of Catalonia, Colom 11, 08222 Terrassa (Barcelona), Spain*
[3]*Kansai Photon Science Institute, National Institutes for Quantum Science and Technology (QST), Kizugawa, Kyoto 619–0215, Japan*
[4]*Graduate School of Advanced Science and Engineering, Hiroshima University, 1–4–1 Kagamiyama, Higashi–Hiroshima, Hiroshima 739–8527, Japan*
*\*jozsef.seres@tuwien.ac.at*

**Abstract:** We report the realization and characterization of parametric amplification in high harmonic generation around 100 eV using He gas in a double gas jet arrangement. The delay of the seed XUV pulse with respect to the amplifier gas jet was scanned by changing the distance between the gas jets. Experiments and numerical calculations show that parametric amplification occurs within a temporal window of several optical cycles. Strong correlation between the seed and amplifier was observed in a shorter, few optical cycles delay range, which appeared as a modulation of the XUV intensity with an unexpected one optical cycle periodicity instead of half optical cycle. Simulations revealed that the strong correlation and also the unusual periodicity was the consequence of the nonlinear effect produced by plasma dispersion on the parametric amplification process during propagation in the amplifier jet.

## 1. Introduction

To understand the underlying processes in high harmonic generation (HHG), experimental and theoretical studies were performed in the last decades. A type of parametric interaction, the x-ray parametric amplification (XPA), which takes part in the HHG process has been recognized [1, 2]. XPA was demonstrated in Ar gas at around 40 eV photon energies [1, 3, 4] and in He gas the process was observed from around 100 eV [5, 6, 7] to higher photon energies [8] up to keV [9, 10]. The underlying process was described from different viewpoints; classical, semi-classical or quantum mechanical. It was considered a high-order parametric amplification [1, 11-14] also accompanied with parametric resonance [1, 13]. It was further understood as stimulated recombination or as a scattering of the electrons on atoms [1, 2, 13, 15, 16]. Besides, it was described as transient gain by populating excited states in the continuum [3, 7, 17, 18].

Using an arrangement of a double-jet sequence instead of a single gas jet, interference between the sources [19, 20], quasi phase matching and parametric gain [1] were demonstrated and essential improvement of the generation efficiency was reached. XPA provides an XUV beam with much smaller divergence than conventional HHG beams have [5].

The generated ultrashort extreme ultraviolet (XUV) amplified pulses are attractive tools to study ultrafast processes in gases and solids. Being well synchronized to their generating laser pulse, they can support pump-probe experiments even in the sub-fs time scale with low temporal jitter. Also, the low divergence of the beam can be suitable to directly seed plasma X-ray lasers [4, 21-23] without using lossy XUV mirrors. For instance, by slightly tuning the fundamental wavelength of 800 nm from a Ti:sapphire laser, the 59[th] order harmonics energy at 91.8 eV (13.5 nm) can be shifted to be a suitable seed pulse for Ni-like silver plasma X-ray lasers (89.2 eV, 13.9 nm) to provide fully spatially and temporally coherent bright beams for demanding applications [21, 24].

In the present study, we examine the optimal conditions for double-jet arrangements to get an intense and low divergent XUV beam at around 100 eV photon energies.

## 2. Experimental setup

In the experiments, in order to measure the delay dependent gain in He, a Ti:sapphire laser system was used. Pulses with parameters of 800 nm central wavelength; 80 fs duration; 30 mJ energy; 10 Hz repetition rate and 20-mm beam diameter were focused by a lens with focal length of 4000 mm to a double-gas jet arrangement, as shown in Fig. 1. In the focus, the beam waist was ≈200 μm and the laser intensity was up to 1 PW/cm$^2$. The gas jets were molybdenum tubes to avoid laser damage. 1 mm holes for gas supply was drilled into the tubes with 3 mm diameter as can be seen in the inset of Fig. 1. The 40-mm-long Rayleigh length of the focused beam was much smaller than the 3 mm length of the gas jets and even longer than the 25 mm maximal distance of the jets used in the measurements. The first He gas jet (Jet1) produced the seed pulse for the second amplifier He jet (Jet2). Helium gas was supplied by a fast solenoid valve to minimize the gas flow and background pressure in the vacuum system to suppress the XUV absorption of the ambient gas. The jets were mounted on motorized translation stages to precisely set the jet positions along the focused laser beam and to set the distance between them. The HHG beam was monitored by an x-ray CCD camera (back illuminated, 13 μm pixel, 1-inch squared detection area) for beam spatial profile measurements and after that, spectral measurements were made by an XUV spectrometer. The spectrometer used was a grazing incidence type and had a 1200-grooves/mm flatfield grating. A spherical mirror was also installed in front of the entrance slit to measure HH beam divergences.

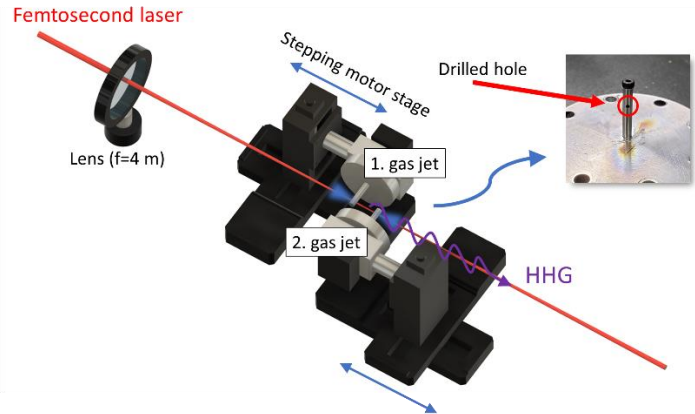

Fig. 1: Experimental setup shows the two-jet sequence used for the experiments. Their position and distance were adjustable. The inset shows the photo of the Mo tube used as a gas nozzle with a hole drilled in for gas outlet and for the laser beam passing through.

## 3. Experimental results

We are interested in the behavior of the 59$^{th}$ harmonic line because it lies near 13.5 nm (91.9 eV) to be usable for coherent lithographic and diagnostic applications, seeding x-ray lasers and free electron lasers. With our experimental conditions (in this case: 0.2 bar He gas in Jet1; 0.25 bar He gas in Jet2 and jet distance of 10 mm), the 59$^{th}$ harmonic line can be very well produced. The intensity of the harmonic line is in the saturation regime, as it can be seen in Fig.2(a), making the generation of the line robust and non-sensitive to the energy fluctuation of the pump laser. Additionally, the wavelength of the harmonic line remains almost intact below saturation, as seen in Fig. 2(b), and only a small blue shift of the line can be observed at beyond-saturation pulse energy as a consequence of the strong ionization of the gas. Beyond phase matching effects, the influence of XPA can be well observed even in the saturation regime. Indeed, the intensities of the harmonic lines are much stronger, as it can be seen in Fig. 2(d), than the intensities that can be expected from constructive superposition of the weak seed (Jet1) and the amplifier (Jet2) spectral lines generated independently in the two jets, which are plotted in Fig. 2(c). The effect is especially strong near 110 eV. Here, the intensities of the harmonic lines from the independent jets are below 100 (counts) while they can be over 1000 when the two jets operate together. Even a stronger XPA effect was observed in He gas using shorter laser pulses [13], where the harmonic lines about 110 eV were observed stronger than the lines below 100 eV. Further properties of the XPA process, like beam profile etc. will be discussed below.

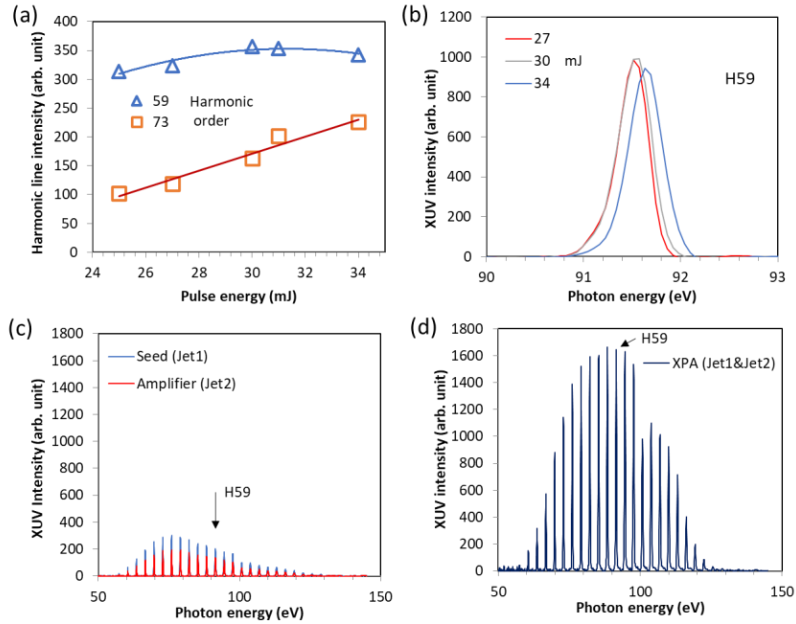

Fig. 2: (a) The harmonic line of H59 is saturated under experimental conditions of the appropriate energy of the pump laser pulse, while near the cutoff, the intensity of the harmonic line H73 increases. (b) At the saturated regime, the line shape of the harmonic H59 remains almost intact and only a slight blue shift can be observed at larger pump energy beyond saturation. (c) The two gas jets alone generated weak harmonic spectra. (d) When both jets operated simultaneously, the intensities of the generated harmonic lines were much stronger than the coherent constructive superposition of lines from the two jets operating separately, showing the presence of the XPA process.

For optimization of the harmonic source and for understanding the underlaying processes, some measurement series were performed where the distance of the gas jets was scanned to create a time delay [19, 5] between the seed XUV pulse produced in Jet1 and the gain medium of Jet2. The measurements used 30 mJ pulse energy from the pump laser what was the optimal for generating the 59$^{th}$ harmonic line according to Fig. 2(a) and corresponded to a peak intensity of about 0.5 PW/cm$^2$ in the focal position and a cutoff harmonic order of 77 (~120 eV). In the first measurement series, 0.2 bar He gas supply was used for Jet1 and its position along the laser beam was set for maximal harmonic intensity. Then, the position of Jet2 (He, 0.25 bar) was scanned within a jet-distance range between 0 and 21 mm. The actual arrangement of the jets and laser beam is plotted in Fig. 3(a) and the HHG spectra were measured at several jet distances. The measurements presented in Fig. 3(b) (left-top panel) were made at a small distance of 2 mm between the jets. In this case, the divergence of the harmonics was 0.7 mrad, which was similar to the value of the harmonic beams generated separately from the individual jets. After reaching a suitable distance of 10 mm, a low divergent beam-part appeared with a divergence of 0.18 mrad. The measured spectra are shown in Fig. 3(b) (left panel). The beam profiles at both jet distances are presented in the right panel of Fig. 3(b) as well as a two-dimensional image for $d = 10$ mm of H59.

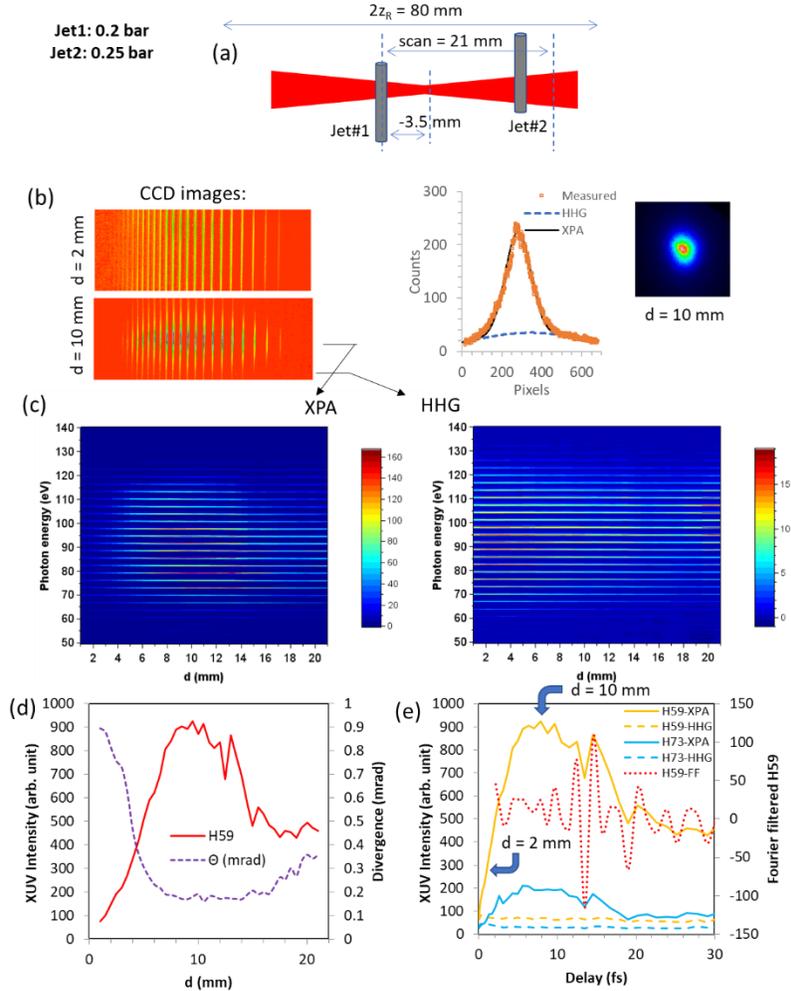

Fig. 3: Experimental results when He pressures of 0.2 bar and 0.25 bar were used to supply Jet1 and Jet2, respectively. (a) Arrangement of the gas jets used in the measurements. Jet2 was scanned within a 21 mm distance range from Jet1. (b) CCD images of spectra were recorded at jet distances when HHG ($d = 2$ mm) and when XPA ($d = 10$ mm) were the dominant contributors. The measured intensity curve of H59 is composed of HHG and XPA beam parts with large and small divergences, respectively. A two-dimensional CCD image of the XPA beam profile was directly measured at around 13 nm after two Zr filters and a Mo/Si multilayer mirror. (c) The intensity of the XPA beam peaked at around 10 mm jet distance, while the intensity of the HHG beam decreased slowly by the jet distance. Note the different intensity scales. (d) Intensity and beam diverge of H59. (e) Intensity curves of H59 and H73 from panel (c) with the calculated delay axis and using solid lines for XPA and dashed lines for HHG. The dotted line plots the Fourier filtered periodic structure appeared on the H59 XPA curve.

The low divergent part of the generated beam (associated to XPA) reacts differently to the change of the jet distance compared to the high divergent part (associated to conventional HHG). We evaluated the signal at the center and outside of the low divergent beam and plotted it in Fig. 3(c). While, for the XPA beam, a well-defined maximum appears between 8- and 10-mm jet distances, the intensity of the HHG beam only shows a slight change by varying the jet distance. At the same time, the beam divergence of H59 sharply decreases at the optimal amplification delays, as plotted in Fig. 3(d), what is the sign of the appearance of the XPA contribution in the generated beam.

In Fig. 3(e), we examine the H59 and H73 harmonics separately, plotting the XPA beam with solid lines and the HHG beam with dashed lines, and a delay scale is assigned to the jet distances [5]. To convert the jet distances to time delays, the difference of Gouy phase and atomic phase between the jets was calculated, considering that Jet2 was scanned and Jet1 was fixed. The delay derived from the jet distance $d$ reads

$$\omega\tau = q\left[\arctan\left(\frac{z_1+d}{z_R}\right) + \frac{a\cdot z_R^2}{z_R^2+(z_1+d)^2}\right] + \varphi_0, \qquad (1)$$

where $q$ is the harmonic order, $z_1$ is the position of Jet1, $z_R$ is the Rayleigh length, $a = \frac{z_R^2+z_1^2}{2z_1 z_R}$, and $\varphi_0$ is determined to set the delay at $d = 0$.

At zero delay (zero jet distance), where there is no XPA component, the two curves (solid and dashed lines in Fig. 3(e)) start from the same value. At larger delays, the HHG beam shows a slight decrease and the intensity of the XPA beam peaks at around 6-10 fs delay range. Beyond the smooth change in the intensity of the XPA beam, a periodic modulation in the H59 intensity curve envelope in Fig. 3(e) can be observed with a period of one optical cycle. To make it more visible, a Fourier filter was applied, and the filtered periodic structure is plotted in Fig. 3(e) with a dotted line. We will discuss the origin and features of the different observed structures in the next section and show how they can be associated to the nonlinear propagation effect of plasma dispersion on the parametric amplification.

For a deeper understanding, for comparison and for obtaining the best conditions to generate a low divergent and bright XUV beam, the same experiment was performed but with higher He gas pressure of 0.3 bar in Jet1. For Jet2 the corresponding optimal pressure was found to be 0.5 bar. The results are plotted in Fig. 4. Similar features as at lower pressures can be observed with a few differences:

(i) At higher gas pressure, the optimal position of Jet1 was at somewhat different position, 9 mm before the laser focus, consequently we made a longer scan range of 25 mm, as shown in Fig. 4(a).

(ii) The contribution of the low divergent XPA component also appeared at certain jet distances. At higher pressure, however, the change of divergence was more characteristic as it can be seen in Fig. 4(d) for H59. The divergence increases somewhat at smaller distances and the XPA beam appears with a low divergence of about 0.11 mrad. The beam profile also can be seen in Fig. 4(b).

(iii) At jet distances between 15 mm and 20 mm, the XPA contribution had a well-defined maximum, while the HHG contribution hardly changed, increasing only slightly as it can be seen in Fig. 4(c).

(iv) The periodic structure is also present in the intensity curve envelope as plotted in Fig. 4(e). This periodic structure is more prominent at higher pressure and its relative position to the intensity curve envelope shifted to the direction of the leading edge from the falling edge. Its absolute position was also shifted from 14 fs to 10 fs delay.

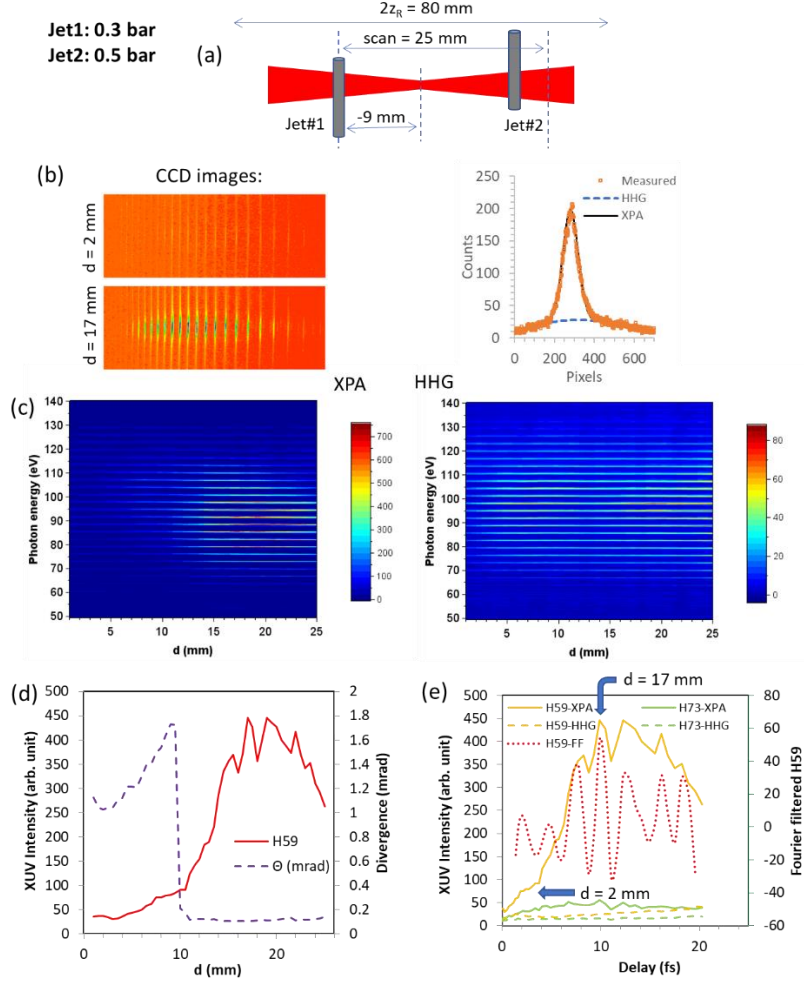

Fig. 4: Similar figures as in Fig. 3 but with different He pressures of 0.3 bar and 0.5 bar in Jet1 and Jet2.

## 4. Discussion

The similarities and differences between the measurements at different pressures can be expected and explained easily from previous studies for the (i)-(ii) observations described in the previous section.

(i) At higher gas pressure, the optimal position of Jet1 is indeed expected to be different; it should be at a larger distance from the focus, in order to decrease the free electron density caused by a smaller field intensity, and also another Gouy phase is expected to fulfill the phase matching condition.

(ii) The divergence of the XPA beam is affected by the small-signal gain in Jet2 [5]. The geometry of the two jets is the same, which is also true for the divergence of the HHG beams from the two jets alone: $\Theta_{seed} \approx \Theta_{ampl}$. In that case, the small signal gain can be estimated with the simple formula $g_0 \approx (\theta_{seed}/\theta_{XPA})^2 - 1$. Using the measured divergencies, 0.7 mrad and 0.18 mrad for 0.25 bar, and 0.6 mbar and 0.11 mbar for 0.5 bar, one obtains the small signal gains to be 14 and 29 in the two cases, respectively. Therefore, as it is expected, the small-signal gain is proportional to the gas pressure in the amplifier jet.

The observations (iii) and (iv) need a particular attention. To understand which temporal delay structure appears and how it changes by the gas pressure, we performed numerical calculations. The generated HHG pulse train in Jet1 is calculated and used as seed XUV pulse together with the laser pulse in Jet2 considering different delays between them. Beyond the joint interaction of the laser and XUV pulse with the gas medium in Jet2, the propagation of them in the gas is also calculated. The theoretical model is based on the single-atom response calculated by solving the Schrödinger equation in the strong field approximation (SFA) in the nonadiabatic form, so that the full electric field of the laser $E(t)$ pulse is used to calculate the nonlinear dipole moment $x(t)$ [25]

$$x(t) = i\int_0^t dt' \left(\frac{\pi}{\varepsilon + i(t-t')/2}\right)^{3/2} \times$$
$$d^*[p_{st}(t,t') + A(t)]e^{-iS_{st}(t,t')}d[p_{st}(t,t') + A(t')]E(t') + c.c.$$

(2)

where

$$p_{st}(t,t') = \frac{1}{t-t'}\int_{t'}^{t} dt'' A(t'')$$

(3)

is the stationary value of the canonical momentum and

$$S_{st}(t,t') = I_p(t-t') - \frac{1}{2}p_{st}^2(t,t')(t-t') + \frac{1}{2}\int_{t'}^{t} dt'' A^2(t'')$$

(4)

is the stationary value of the action. $A(t)$ is the vector potential of the laser field, which is considered linearly polarized, $I_p$ is the atomic ionization potential and $\varepsilon$ is an infinitesimal constant. We consider the case of hydrogen-like atoms, for which the dipole matrix element for transitions to and from the continuum with momentum $k$ can be approximated as

$$d(k) = i\frac{2^{7/2}(2I_p)^{5/4}}{\pi}\frac{k}{(k^2 + 2I_p)^3}.$$

(5)

We assume ground-state depletion by using the tunnel ionization rate in the ADK theory [26], and the single-atom HHG power spectrum is calculated from the Fourier transform of the acceleration of the dipole in Eq. (2). We consider a driving laser field composed of a Gaussian temporal profile 26 fs (FWHM) 800 nm IR strong pulse of $10^{15}$ W/cm$^2$ peak intensity. Since the simulations are very computationally expensive a duration of the IR pulse somewhat shorter than in the experiments is considered. As commented above, the seed field is produced by a single-atom interaction from the intense IR pulse in He at the Jet1. This seed pulse combined at different delays with the intense IR pulse is used as input for the interaction with a first numerical cell of He atoms in the second Jet2. The HHG output from this first interaction in the Jet2 together with the seed and IR pulses are propagated and used as input for a second interaction with a second cell of He atoms. The process is repeated iteratively, so that propagation is described in 1D in a particle-in-cell simulation scheme. In order to fully consider the macroscopic effects associated to propagation, we take into account the regular phase mismatch associated to neutral gas and dispersion from the free electrons together with the geometrical phase mismatch due to the shape of the driving laser pulse, which arises primarily from the Gouy phase shift due to the focused laser beam. Neutral gas dispersion and absorption are calculated from the scattering cross sections ($f_1$ and $f_2$) with data obtained from Ref. [27].

The use of a 1D propagation model is adequate in the present case because essential spatial distortion of the laser pulse passing thorough the jets was not observed in the experiments.

Therefore, only the pressure (gas density) - length product is relevant. In order to optimize the computational resources, we hence used 40-times higher pressures (10 bar and 20 bar instead of 0.25 bar and 0.5 bar) and a proportionally shorter propagation length of 75 µm. The small-signal gain and XPA are not affected either by this scaling because they are also proportional to the gas density – length product. It was found that an about 2-times longer propagation length was necessary to reproduce the observations of the experiments, which shows that the ionization rate of the gas and so the nonlinear refractive index was taken somewhat smaller in the simulations than the actual ionization rate experienced by the He gas. The numerical modelling shall allow us to understand the experimental observations and to analyze the optimal conditions for the production of brighter, more efficient coherent XUV beams near 13 nm. The calculation results are presented in Fig. 5(a) and 5(b) and compared with the measurements.

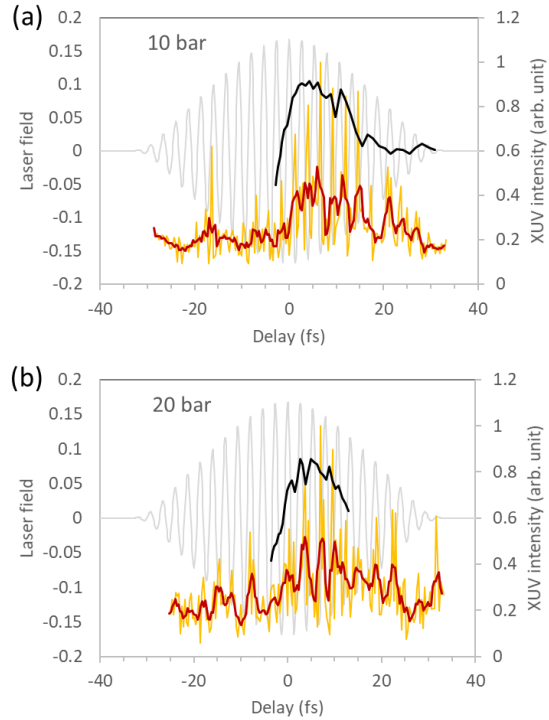

Fig. 5: Comparison of the numerical calculations (orange and red lines) to the measurements (black lines) for two He gas pressures of 10 bar (a) and (b) 20 bar. The calculations were made using a 26 fs long laser pulse. The red lines are the averaged calculations over 1.3 fs range.

The calculation results (orange lines) show well-defined, high contrast peak structure in the 0 to 20 fs delay range with the observed periodicity of one optical cycle. The results from the simulations look different from the experimental results, where only a relatively small modulation appears over a larger smooth background. However, after making an averaging of the calculation over 400 nm delay range (red lines), the results reproduce all the features of the measurements reasonably well. In Fig. 5(a), - there is a fast increase of the signal at 0 delay; - a small intensity modulation up to 10 fs; - followed by a short, large intensity modulation with a big hole at about 10 fs in the falling edge; - and the signal decrease to a smaller value. The differences between the two pressures are also reproduced by the calculation. In Fig. 5(b), - after the fast increase of the signal at 0 delay; - a large intensity modulation appears up to 10 fs delay, where the smooth background has a maximum; - and the signal decreases after that.

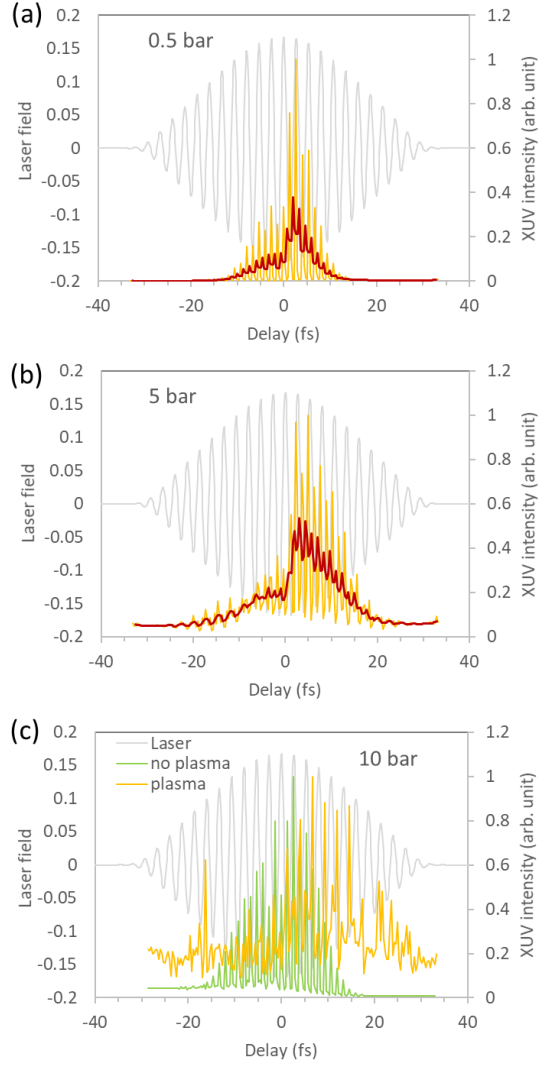

Fig. 6: (a) Low pressure (0.5 bar) calculations to eliminate the nonlinear propagation effects. The modulation periodicity is clearly half optical cycle. (b) Calculations at an intermediate pressure of 5 bar, where nonlinear effects start to appear, although the half-cycle periodicity is still visible. The red lines are the averaged calculations over 1.3 fs range. (c) At 10 bar, the modulation periodicity becomes one optical cycle. However, if the calculations are performed without taking the plasma dispersion into account, the half-optical periodicity is restored (green line).

The measurements and the corresponding calculations suggest that the difference between the delay curves at the two different pressures might originate from the nonlinear propagation effects in Jet2 added to the anyway present x-ray parametric amplification. Indeed, x-ray parametric amplification alone should result in a structure with half optical cycle [2], contrary to the observed and calculated one optical cycle. However, the nonlinear refractive index of the gas and the plasma dispersion of the free electrons, which are nonlinear to the laser intensity, can cause a distortion of the laser field during propagation that ultimately affects the parametric amplification process. Both contributions are proportional to the gas density, so that to essentially decrease the nonlinear contribution, the calculations were repeated at a more than one order of magnitude lower pressure of 0.5 bar, which is plotted in Fig. 6(a). As it can be seen, a periodic structure with half optical cycle is clearly produced. At an intermediate pressure of 5 bar, Fig. 6(b), the calculations still produce the half optical cycle periodicity, but the effect of the nonlinearity starts to appear and the amplitudes of every second peaks get smaller. Applying the same averaging over 1.3 fs delay window as in Fig. 5, the half-optical cycle periodicity can still be observed in Fig. 6(b). To further address the change of the periodicity observed, the simulations were performed by neglecting different contributions as the Gouy phase, the atomic dispersion, and the plasma (free electron) dispersion separately, in the case of propagation for 10 bar. The Gouy phase and atomic dispersion made no or minimal effect, however the lack of plasma dispersion altered the results dramatically as can be seen in Fig. 6(c). Without plasma dispersion, the delay curve has a periodicity with half optical cycle and the position of the maximum is also shifted. The simulations hence reveal that the unusual periodicity and precise delay region for optimal amplification can be associated to the nonlinear propagation effect produced by plasma dispersion on the parametric amplification process.

Additionally, in order to address the necessary averaging performed in the simulations results, we checked the effect of carrier envelope phase fluctuation and energy fluctuation of the laser. We concluded that both have no real effects in the structure since the delay positions were preserved. Another possible reason for the averaging need would be the distance uncertainty/fluctuation between the two gas jets. The measurements were recorded over 30 s, meaning the integration of 300 laser shots. The small distance uncertainty between shots can be the result of the mechanical vibration of the gas jets caused by the switching valve and/or the push of resulted gas pulses. However, this should have been caused a large shot-to-shot fluctuation of the XUV intensity, which was not observed. We are left with the 1D propagation considered in the simulations, which we hence associate to the averaging need, since the phase fronts of the focused Gaussian laser beam are somehow curved at the positions of the gas jets, meaning that different optical lengths between the gas jets within the diffraction angle of the laser beam may be non-negligible.

## 5. Conclusion

We have examined both experimentally and with numerical simulations the optimal conditions for double-jet arrangements to get an intense and low divergent XUV beam by XPA. We have observed one optical cycle periodicity on the envelope on the delay intensity curves of XPA measurements and have shown how the relative position of the periodic structure to the intensity envelope curve changes with the applied pressure in the gas jets. The differences observed in the measurements can be understood as the effect of the nonlinear propagation in the gas, namely the intensity dependent free electron density that affects the field of the laser pulse propagating in the gas and consequently the x-ray parametric amplification process.

We are interested in the behavior of the 59$^{th}$ harmonic line because it lies near 13.5 nm (91.8 eV) being suitable for coherent lithography. To boost the pulse energy, the harmonic can be used for seeding a plasma lasing medium for further amplification. For best fitting the wavelength, $Li^{2+}$, $C^{5+}$ or $Xe^{10+}$ x-ray lasers can be considered, but the population inversion in these lasing media is created by means of a recombination scheme. In principle, the recombination x-ray laser has a long pulse duration above nanosecond, and the seed high harmonic pulse duration is too short to obtain the intense x-ray laser pulse. Near this wavelength, at 13.9 nm, a Ni-like Ag ion (4d-4p) plasma x-ray laser is the best candidate for a seed-amplification scheme, because the population inversion is created by means of a transient collisional excitation scheme. Such x-ray laser has a pulse duration of less than several picoseconds and, consequently, the temporal matching between the seed high harmonic and the x-ray laser is much better. With this type of x-ray laser, lithographic applications [28] were successfully demonstrated. In addition, the 13.5-nm high harmonic beam is expected to be applicable as a seeding beam of spatial coherent EUV free electron lasers (FEL), by which fully coherent (temporally and spatially) EUV FEL can be obtained. Bright 13.5-nm coherent pulses provide powerful diagnostic/developing tools for EUV lithography, such as inspecting a defect and developing a new EUV photoresist.

**Funding**

Ministerio de Economía y Competitividad (FIS2017-85526-R); JSPS KAKENHI JP20H00141, JP19K15402, JP21H03750; JKA and its promotion funds from KEIRIN RACE; Horizon 2020 Framework Programme (856415); European Metrology Program for Innovation and Research (20FUN01 TSCAC); Österreichische Nationalstiftung für Forschung (AQUnet)

**Acknowledgments**

Computation time was provided by the Supercomputer System, Institute for Chemical Research, Kyoto University.

**Disclosures**

The authors declare no conflicts of interest

**Data availability statement**

Data underlying the results presented in this paper are not publicly available at this time but may be obtained from the authors upon reasonable request.